\documentstyle[epsfig]{aipproc}

\def\stackunder#1#2{\mathrel{\mathop{#2}\limits_{#1}}}%

\begin{document}

\title{Non-local regularization of chiral quark models in the soliton sector%
\thanks{%
talk presented by G. Ripka at the
International Workshop on Hadron Physics ``Effective Theories of Low Energy
QCD'', Coimbra, Portugal, September 10-15, 1999
}}
\author{Georges Ripka$^*$ and Bojan Golli$^{\dagger}$}
\address{$^*$Service de Physique Th\'{e}orique,
Centre d'Etudes de Saclay\\
F-91191\ Gif-sur-Yvette Cedex, France%
\thanks{E-mail: ripka@spht.saclay.cea.fr}\\
$^{\dagger}$Faculty of Education, University of Ljubljana and
J.Stefan Institute,\\
Jamova 39, P.O.Box 3000,
1001\ Ljubljana, Slovenia%
\thanks{E-mail: Bojan.Golli@ijs.si}}

\maketitle
\begin{abstract}
A chiral quark model is described which is regularized in terms of 
Lorentz invariant non-local interactions. The model is regularized to 
all loop orders and it ensures the proper quantization of the baryon 
number. It sustains bound hedgehog solitons which, after suitable 
centre of mass corrections, can adequately describe the nucleon.
\end{abstract}

\section{Some specificities of chiral quark models}

This work was done in collaboration with Wojciech Broniowski from Krakow.
We consider chiral quark models which encompass three sectors. The vacuum 
and soliton sectors, which are treated in the mean-field (leading order in 
$N_c$) approximation, and the meson sector, which describes the (next to 
leading order in $N_c$) vibrations of the vacuum sector. Not all models are
applicable to the three sectors. For example, constituent quark models, in
which quarks interact with confining forces, cannot describe the vacuum
sector, that is, the Dirac sea. However, they can and do describe the
excited states of baryons, a thing which the chiral quark models cannot do
(except, possibly, the $\Delta $) for lack of confinement.

Chiral quark models (nor any of the other low energy quark models) have not
been derived from QCD. The only serious attempt to derive them from QCD\ is
the instanton gas model \cite{Diakonov86,Shuryak82}. In this approach, the
chiral quark model is derived by calculating the propagation of quarks
in a gas of instantons. A regularized effective theory results,
as it should. It predicts both the value of the cut-off and the
form of the regulator. The non-local regularization discussed here has the
same form as the one derived from the instanton gas model. Unfortunately,
the quark models derived from the instanton structure of the vacuum do not
lead to quark or color confinement. This serious limitation serves as a
reminder that we have not really succeeded in deriving low energy effective
theories from QCD.

Other so-called ``derivations'' of quark models from QCD involve more
guesswork than derivation. Most telling is their inability to derive a
regularized model. If infinities appear in an effective theory, one should
seek the physical processes which prevent the infinities from occurring.
Invoking the roughly $200$~MeV QCD cut-off is not a serious argument. Nor
does QCD\ imply in any sense that the quark-quark interaction at low energy
should be a one-gluon exchange with a modified gluon propagator. The
regularizations used so far in the Nambu Jona-Lasinio type models for
example (proper-time regularization being the most commonly used one so
far), are nothing but renormalization techniques in which a finite cut-off
is maintained. Not only is this arbitrary but such regularizations are
flawed with problems.

One might argue that the value of the cut-off should not matter. Indeed it
would not if the effective theory consisted, for example, in eliminating
some high energy degrees of freedom and using the remaining degrees of
freedom to work out the dynamics of low energy phenomena. In such a case,
one might expect the cut-off to be much larger than the inverse size of the
composite particles and the results not to be sensitive to the cut-off. In
chiral quark models, however, this is not the case. The cut-offs required
to fit $f_\pi $ are about $700$~MeV, hardly larger than the $\rho $ or the
nucleon mass. This is a fact of life, whether we like it or not. One can of
course simply discard such models, but better models do not seem to be
forthcoming.

\section{The soliton in the non-local chiral quark model}

The non-local chiral quark model is defined by the euclidean action:
\begin{equation}
  I\left( q,q^{\dagger }\right) =
 \left\langle q\left| \partial _\tau +\frac{\vec{\alpha}.\vec{\nabla}}i
     +m\right| q\right\rangle 
  -\frac{G^2}2\int d_4x\left( \left\langle q\left| r\right| x\right\rangle 
  \beta \Gamma_a\left\langle x\left| r\right| q\right\rangle \right)^2 \;. 
\label{nonlocact}
\end{equation}
In this expression, $\Gamma _a=\left( 1,i\gamma _5\tau _a\right) $, 
$q\left(x\right) \equiv \left\langle x\left| q\right. \right\rangle $ 
is the quark field, and $r$ is a regulator. 
The regulator is assumed to be diagonal in
momentum space and it has a range which defines an effective euclidean
cut-off $\Lambda $. For example, we could take 
$\left\langle k\left|r\right| k^{\prime }\right\rangle =
\delta _{k,k^{\prime }}r\left( k^2\right)$ with 
$r\left( k^2\right) =e^{-\frac{k^2}{2\Lambda ^2}}$, where $k$ is a
euclidean 4-vector $k_\mu =\left( \omega ,\vec{k}\right) $ with 
$k^2=\omega^2+\vec{k}^2$. 
The interaction term of the action (\ref{nonlocact}) can be viewed as a 
contact 4-fermion interaction involving the \emph{delocalized
quark fields}:
\begin{equation}
 \psi \left( x\right) =
 \left\langle x\left| r\right| q\right\rangle =
 \int d_4y\,\left\langle x\left|r\right|y\right\rangle \,q\left(y\right)\;.
\label{deloc}
\end{equation}

An action of the form (\ref{nonlocact}) is derived from the instanton gas
model of the QCD vacuum \cite{Diakonov86,Shuryak82}, which predicts a
cut-off function of the form:
\begin{equation}
  r\left( k^2\right) =f\left( k\rho /2\right)\;, 
\quad \quad \quad 
  f\left(z\right) = -z\frac d{dz}\left(I_0\left(z\right) K_0\left(z\right)
                    -I_1\left(z\right) K_1\left(z\right)\right)  
\label{instanton}
\end{equation}
where $\rho $ is the instanton size. The the cut-off is determined by the
inverse instanton size $\rho$. The form (\ref{instanton}) has 
$r\left(z=0\right) =1$ and $r\left( z\right) 
\stackunder{z\rightarrow \infty}{\rightarrow }\frac 9{2k^6\rho ^6}$. 
However, at large euclidean momenta $k$,
the form (\ref{instanton}) is no longer valid and the cut-off function is
dominated by one gluon exchange. It decreases as $\frac 1{k^2}$ (with
possible logarithmic corrections) and not as $\frac 1{k^6}$. We find that
the fall-off of the regulator at large euclidean $k^2$ does not affect
the soliton properties very much. For this reason, we have felt free to use
various simple forms of cut-off functions, such as a gaussian, which have an
additional advantage in that they can be analytically (although arbitrarily)
continued to negative values of $k^2$. We shall see below that the analytic
continuation is required to include the valence orbit. Similar
regularization has been used by the Manchester group \cite{Birse98} in the
meson and vacuum sectors. Various regularization schemes are reviewed in
chapter 6 of Ref.\cite{Ripka97}.

The euclidean action allows us to calculate the partition function 
$Z=\int D\left( a\right) D\left( a^{\dagger }\right) 
e^{-I\left( a,a^{\dagger}\right) }$ and the ground state energy 
$E=-\frac \partial {\partial \beta}\ln Z.$ 
The partition function cannot be written in the form $Z=Tr\,e^{-\beta H}$ 
because the regulator in the action (\ref{nonlocact})
prevents us from defining a hamiltonian $H$. We are also unable to quantize
the quark fields but we shall see that the baryon number is nonetheless
properly quantized.

We work with the equivalent bosonized form of the action:
\begin{equation}
  I\left( \varphi \right) =
  -Tr\ln \left( \partial _\tau +\frac{\vec{\alpha}.\vec{\nabla}}i+\beta m
   +\beta r\varphi _a\Gamma _ar\right) 
   +\frac 1{2G^2}\int d_4x\,\varphi _a^2\left( x\right)  
\label{isvs}
\end{equation}
in which case the partition function is given by the path integral 
$Z=\,\int D\left( \varphi \right) e^{-I\left( \varphi \right) }$. We 
refer to $\varphi _a\Gamma _a=S+i\gamma _5\tau _aP_a$, as the ``chiral 
field'' and we say that the chiral field is ``on the chiral circle'' if, 
for all $x$, we have $S^2\left( x\right) +P_a^2\left( x\right) =M_0^2$, 
where $M_0$ is an $x$-independent constant mass.

We have calculated a localized and time independent stationary point of the
action (\ref{isvs}), consisting of a chiral field with a hedgehog shape 
$S\left(r\right) +i\gamma _5\widehat{x}_a\widehat{\tau }_a P\left(r\right)$ 
\cite{Ripka98}. The shape of the fields and the soliton energy can be
calculated in terms of the energies $e_\lambda \left( \omega \right) $ of
the quark orbit. The ``Dirac hamiltonian'' is diagonal in the energy
representation, although it remains energy dependent. The quark orbits 
$\left| \omega ,\lambda _\omega \right\rangle $ satisfy the equations :
\begin{equation}
  \partial_\tau \left| \omega ,\lambda _\omega \right\rangle =
 i\omega \left|\omega ,\lambda _\omega \right\rangle \;,
\quad \quad \quad 
  \left(\frac{\vec{\alpha}.\vec{\nabla}}i+\beta m
  +\beta r\varphi_a\Gamma_ar\right)
  \left| \omega ,\lambda _\omega \right\rangle =
  e_\lambda \left(\omega\right) \left| \omega ,\lambda_\omega 
 \right\rangle \;.
\label{basis2}
\end{equation}
The energy of the soliton is:
\begin{equation}
  E_{sol}=N_ce_{val}+\frac 1{2\pi }\int_{-\infty }^\infty \omega d\omega
  \,\sum_{\lambda _\omega }\frac{i+\frac{de_\lambda \left( \omega \right) }
  {d\omega }}{i\omega +e_\lambda \left( \omega \right) }
  +\frac 1{2G^2}\int d_3x\,\varphi _a^2\left( \vec{x}\right) -vac.
\end{equation}
where $-vac.$ means that we subtract the vacuum energy. In the vacuum, 
$P=0, $ $S=M_0$ and there is no valence orbit contribution $e_{val}$. 
The latter is discussed in the next section.

\section{The quantization of the baryon number and the valence orbit}

\label{sec:barnum}

We calculate the baryon number from the Noether current associated to the
gauge transformation 
$q\left(x\right) \rightarrow e^{-i\alpha\left(x\right)}q\left(x\right)$. 
It turns out to be:
\begin{equation}
  B=-\frac 1{2\pi iN_c}\int_{-\infty }^\infty d\omega \,
  \sum_{\lambda _\omega}\frac{i+\frac{de_\lambda \left( \omega \right) }
   {d\omega }}{i\omega+e_\lambda \left( \omega \right) }\;.  
\label{number}
\end{equation}
The extra term $\frac{de_\lambda \left( \omega \right) }{d\omega }$ in the
numerator arises from the fact that the regulator $r$ does not commute with 
$\alpha \left( x\right) .$ Its effect is to make the residues of all the
poles of the quark propagator $\frac 1{i\omega +e_\lambda \left( \omega
\right) }$ equal to unity. This effectively quantizes the baryon number 
in a manner which does not seem to be related to the topology of the 
hedgehog field.\footnote{%
Nor is the soliton stabilized by the topology of the chiral field.} 
This is most fortunate because, a priori, there is no reason to expect a 
theory, in which we cannot quantize the quark field, to yield a properly 
quantized baryon number.

The expression (\ref{number}) suggests a way to include the valence orbit so
as to ensure that the baryon number of the soliton, relative to the
vacuum, is equal to unity. We
calculate ``on-shell'' pole of the quark propagator in the hedgehog
background field by searching for a solution of the equation $\left. i\omega
+e_\lambda \left( \omega \right) \right| _{\omega =ie_{val}}=0$. Because of
the regulator, the solutions are scattered all over the complex $\omega $
plane. However, it is well known that, in the local theory, where we set 
$r=1 $, and for a hedgehog field with winding number unity, a well separated
bound orbit with grand spin and parity $0^{+}$ occurs with energy $e_{val}$
close to zero \cite{Ripka84}. In the non-local theory, we find that a
solution of the equation $\omega =ie_{val}\left( \omega \right) $ can always
be found on the imaginary $\omega $ axis, close to the origin $\omega =0$,
and that no other pole occurs in the vicinity. We therefore ensure that the
soliton has a baryon number $B=1$ by deforming the integration path over 
$\omega $ in such a way as to include the contribution of this pole. This
requires an analytic continuation of the regulator. Such a continuation is
arbitrary but the analytic continuation does not extend as far from the
origin as $e_{val}$. Indeed, since the soliton size is small, $\vec{k}^2>0$
is large and this, on the average, makes $k^2=-e_{val}^2+\vec{k}^2$ less
negative. Unfortunately however, the form (\ref{instanton}) of the
regulator, predicted in the instanton model, does not allow any analytic
continuation whatsoever, thereby, strictly, prohibiting its use in the
soliton calculation.

\section{Results of self-consistent soliton calculations}

The model parameters are the coupling constant $G$ appearing in the
lagrangian, the cut-off $\Lambda $ appearing in the regulator and the
current quark mass $m$. The values of the three parameters are constrained
by fitting the pion decay constant $f_\pi =93$~MeV and the pion mass to
$m_\pi =139$~MeV. The expression used to calculate the pion decay
constant $f_\pi $ is:
\begin{equation}
  f_\pi^2=2N_fN_cM_0^2\int \frac{d_4k}{\left( 2\pi \right) ^4}
  \frac{r_k^4-k^2r_k^2\frac{dr_k^2}{dk^2}+
   k^4\left( \frac{dr_k^2}{dk^2}\right) ^2}{\left( k^2+r_k^2M_0^2\right)^2}
\label{fpi}
\end{equation}
valid in the chiral limit $m\rightarrow 0$ and it is not identical to the
Pagels-Stokar formula \cite{Pagels79}. This leaves one undetermined
parameter which we choose to be the constituent quark mass $M_0$ at zero
4-momentum. The pion decay constant $f_\pi $ sets the scale. Grossly,
soliton energies increase and soliton radii diminish as $f_\pi $ increases
(see table \ref{golli5}).

\begin{figure}[h] 
\centerline{\epsfig{file=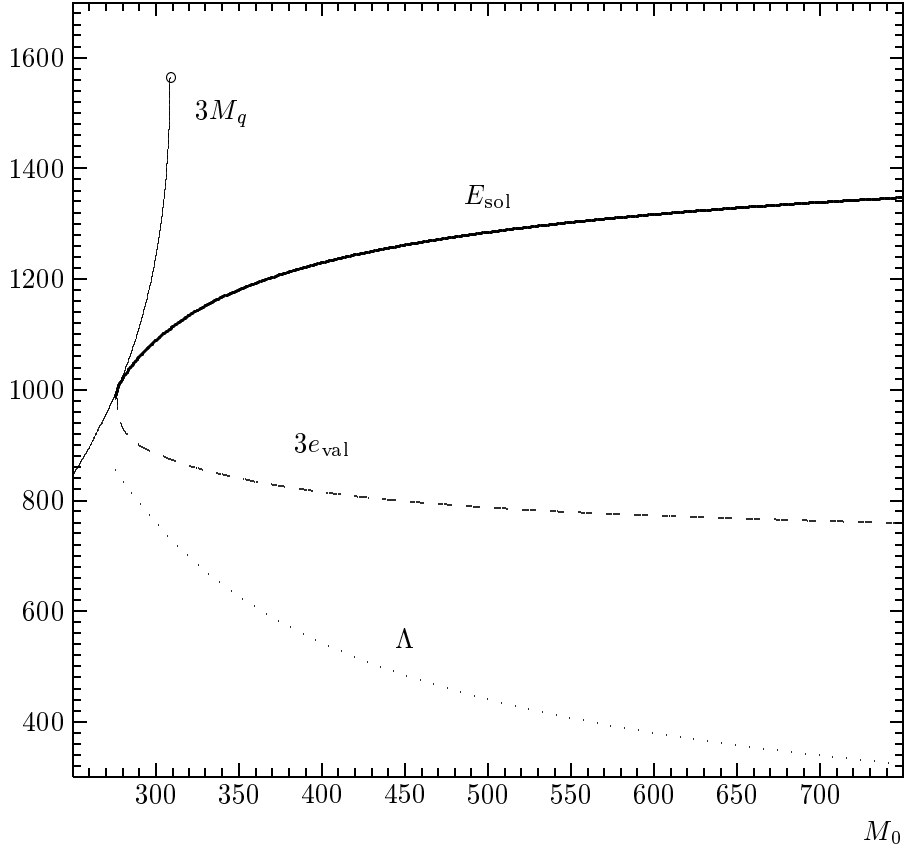,width=120mm}}
\vspace{10pt}
\caption{The energy of the soliton [in MeV] (bold solid line), $N_c$ times
the free-space quark mass (solid line) and the valence contribution
to the soliton energy (dashed line) plotted as functions of the
parameter $M_0$ [in MeV].
A Gaussian regulator is used;
$\Lambda$ (dots) is fitted to $f_\pi=93$~MeV.}
\label{fige}
\end{figure}

Figure \ref{fige} shows the soliton energy $E_{sol}$
as a function of the free
parameter $M_0$. A soliton is a bound state of $N_c=3$ quarks which
polarize the Dirac sea. With a gaussian regulator, it is formed if 
$M_0\gtrsim 276$~MeV, that is, for a sufficiently strong coupling 
constant $G\gtrsim 4.7\times 10^{-3}$~MeV$^{-1}$. 
The bound state occurs when the
energy of the system is lower than the energy $N_cM_q$ of $N_c$ free
constituent quarks in the vacuum: $E_{sol}<N_cM_q$. The mass $M_q$ is the
on-shell constituent quark mass, obtained by searching for the pole of the
quark propagator in the vacuum. It is the solution of the equation $\left.
k^2+\left( r_k^2M_0+m\right) ^2\right| _{k^2=-M_q^2}=0$, which requires an
analytic continuation of the regulator to negative values of $k^2$.
Figure \ref{fige}
also shows $N_cM_q$. At the critical value $M_0\approx 276$~MeV, the
two curves merge. The contribution $N_ce_{val}$ of the
valence orbit is also shown. At the critical value of $M_0$, the energy 
$e_{val}$ of the valence orbit, which is the on-shell mass of a quark
propagating in the hedgehog field, becomes a well distinguished bound orbit.

At $M_0\approx 309$~MeV, the curve displaying $N_cM_q$ on 
figure~\ref{fige} abruptly stops. 
Indeed, for larger values of $M_0$, the poles of the quark
propagator no longer occur for real values of $k^2$. This means that quarks
can no longer materialize on-shell in the vacuum. This feature is discussed
in chapter 6 of Ref.\cite{Ripka97} and it has been considered by several
authors as a sign of quark confinement \cite{Krewald92,Roberts94,Birse95}.
In fact, when a pole of the quark propagator disappears from the real $k^2$
axis, it simply moves into the complex plane. Such poles indicate
instability of the assumed vacuum state against the addition of a single
quark.

However, our calculation shows that, in the background soliton field, the
on-shell valence orbit continues to exist and so does the soliton.
Unfortunately, the regulator also introduces extra unwanted poles in the
propagators of colorless mesons, so that the model does not express color
confinement. Similar unwanted poles occur in proper-time regularization \cite
{Ripka95}. Our ignorance as how to continue propagators in the complex $k^2$
plane reflects our ignorance of the confining mechanism \cite{Stingl90}.

\begin{figure}[h] 
\centerline{\epsfig{file=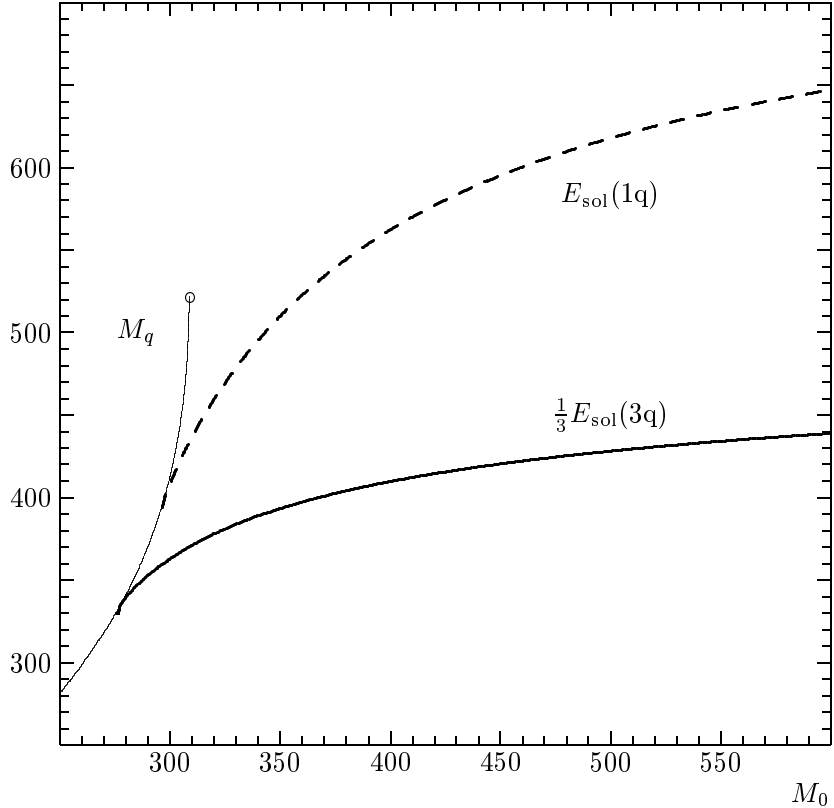,width=110mm}}
\vspace{10pt}
\caption{The energy per quark [in MeV]
for the soliton with three valence quarks (bold line),
the soliton with one valence quark (dashed line)
and the free-space quark mass $M_q$ plotted as functions
of the parameter $M_0$ [in MeV].}
\label{figes}
\end{figure}

Apart from the solitons consisting of three valence quarks
we find stable solitons consisting of a single valence quark
in the background soliton field (see figure \ref{figes}) as well
as of two valence quarks.
Similar solutions have been found in the linear sigma model
with valence quarks \cite{Golli97}.

\begin{figure}[h] 
\centerline{\epsfig{file=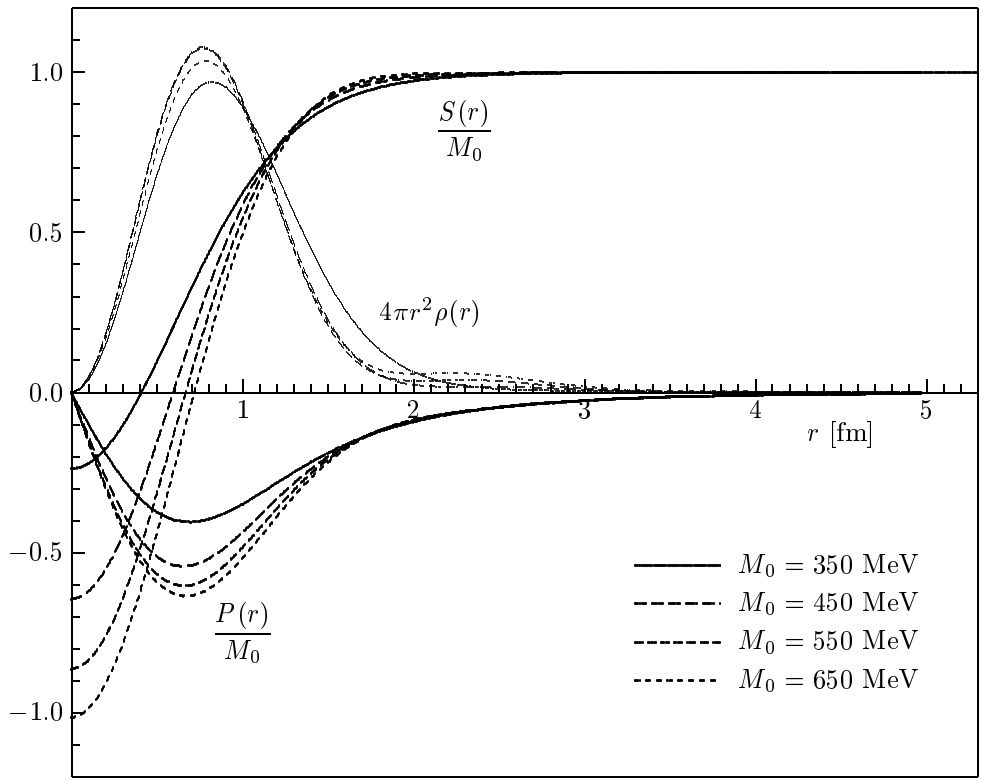,width=130mm}}
\vspace{10pt}
\caption{Self consistently determined fields
and baryon densities ($4\pi r^2\rho$)  for various values of $M_0$;
a gaussian regulator is used.}
\label{figmf}
\end{figure}

Figure \ref{figmf}
shows the scalar and pseudoscalar fields $S\left( x\right) /M_0$
and $P\left( x\right) /M_0$ of the soliton obtained with several values of 
$M_0$, together with the soliton quark density $\rho \left( x\right)$.
Note that, within the soliton, the fields \emph{do not} lie on the chiral
circle and $S^2\left( x\right) +P^2\left( x\right) <M_0^2$. Indeed, the pion
component $P\left( x\right) $ never reaches the values $-M_0$. This is a new
dynamical result. This is the only calculation, as far we know, in which one
can check dynamically whether the chiral field remains or not on the chiral
circle. It could not be checked in the renormalized linear sigma model,
because close lying Landau poles occur which make the soliton unstable
against high gradients in the fields \cite{Ripka87,Perry87}. It could also
not be checked in local theories which use proper-time regularization
because, in such theories, the soliton is unstable unless the fields are
constrained to remain on the chiral circle \cite{Goeke92,Ripka93d}. No such
instability occurs with the non-local regularization.

The soliton we obtain with non-local regularization has a structure which
lies midway between a Friedberg-Lee soliton \cite{Lee81,Wilets89} (in which
the pion field has a vanishing classical value), and a Skyrmion \cite
{Skyrme62,Holzwarth93} (in which the chiral field is constrained to remain
on the chiral circle). This raises the problem of the collective rotational
motion of the soliton. If the deformation in spin and isospin space is
stable enough to sustain a rotation without significant distortion, then the
$\Delta $ can be described as a rotation of the soliton and the $N-\Delta $
mass splitting can be estimated by cranking. If, however, the deformation is
small, the $\Delta $ may be better described as a bound state of quarks with
aligned spins and isospins. We have not tackled this problem yet.

\begin{table}[h] \centering%
\begin{tabular}{|c|c|c|c||c||c|c|c|c|c|}
\hline
$M_0$ & $\Lambda $ & $m$ & $\left\langle \bar{q}q\right\rangle ^{1/3}$ & 
$1/G $ & $e_{\mathrm{val}}$ & $E_{\mathrm{Dirac}}$ & $E_{\mathrm{sol}}$ & 
$\langle r^2\rangle ^{1/2}$ & $g_A$ \\
MeV & MeV & MeV & MeV & MeV & MeV & MeV & MeV & fm &  \\ \hline
300 & 760 & 7.62 & $-215$ & 182 & 295 & 2360 & 1088 & 1.32 & 1.28 \\
350 & 627 & 10.4 & $-200$ & 140 & 280 & 1715 & 1180 & 1.04 & 1.16 \\
400 & 543 & 13.2 & $-185$ & 113 & 272 & 1433 & 1229 & 0.97 & 1.14 \\
450 & 484 & 15.9 & $-173$ & 94 & 266 & 1275 & 1261 & 0.96 & 1.12 \\ \hline
\end{tabular}
\caption{Properties of self-consistent soliton solutions
obtained with a gaussian regulator.\label{golli1}}%
\end{table}%

Table \ref{golli1} shows some properties of calculated solitons for various
values of the mass parameter $M_0$. Rather good values of $g_A$ are
obtained. The soliton mass and energies need to be corrected for spurious
centre of mass motion (see table \ref{golli5}).

The fields which describe the soliton break translational symmetry. The
center of mass of the system is not at rest and it makes a spurious
contribution both to the energy and to the mean square radius (more
generally, to the form factor). This spurious contribution is not measured
and it should be subtracted from the calculated values. The subtraction
occurs at the next to leading order (in $N_c$) approximation. A rough
estimate can be obtained from an oscillator model. If $N_c$ particles of mass
$m$ move in a $1s$ state of a harmonic oscillator of frequency $\hbar\omega$,
the centre of mass of the system is also in a $1s$ state and it contributes 
$\frac 34\hbar \omega =\left\langle P^2\right\rangle/2N_cm$
to the energy. We have therefore corrected the soliton energies by
subtracting $\left\langle P^2\right\rangle/2E_{sol}$ from the
calculated energy. Furthermore, in the oscillator model, the center of mass
contributes a fraction $\frac 1{N_c}$ of the mean square radius, so that we
have corrected the mean square radius by multiplying the calculated value by
a factor equal to $\left( 1-\frac 1{N_c}\right) $.

Table \ref{golli5} shows the result. The soliton energies and radii are then
considerably closer to the experimental values observed in the nucleon.

\begin{table}[h] \centering%
\begin{tabular}{|c|c|c|c|c|}
\hline
$M_0$ & $E_{sol}$ & $\left\langle r^2\right\rangle _{sol}$ & $E_{corr}$ & 
$\left\langle r^2\right\rangle _{corr}$ \\ \hline
\quad MeV\quad & \quad MeV\quad & \quad fm$^2$\quad & \quad MeV\quad & \quad
fm$^2$\quad \\ \hline
300 & 1088 & 1.7 & 965 & 1.1 \\
350 & 1180 & 1.08 & 990 & 0.72 \\
400 & 1229 & 0.94 & 1000 & 0.62 \\
450 & 1261 & 0.92 & 980 & 0.61 \\ \hline
450$^{*}$ & 1458 & 0.69 & 1200 & 0.43 \\ \hline
\end{tabular}
\caption{Elimination of spurious c.m. motion.
Gaussian regulator, $\Lambda$ fitted to $f_\pi=93$~MeV;
$^*$ $\Lambda $ fitted to $f_\pi =1.25\times 93$~MeV.\label{golli5}}%
\end{table}%

\section{Conclusion: why take the trouble?}

The non-local regularization effectively cuts out of the quark propagators
the 4-momenta which are larger than the cut-off. The non-local
regularization makes the theory finite at all loop orders. 
The simpler proper-time and Pauli-Villars regularization schemes regularize 
the quark loop only and they require extra independent cut-offs when next 
to leading order meson loops are included. Both the real and the imaginary 
parts of the action are regularized, while the anomalous properties
remain independent of the cut-off \cite{Cahill88,Holdom89,Ripka93}, and the
baryon number remains properly quantized. In proper time and Pauli-Villars
regularization schemes only the real part of the action is regularized and
the imaginary part is left unregularized in order to enforce correct
anomalous processes. Why not limit the 3-momenta of the quarks, thereby
avoiding unwanted extra poles in the propagators? Breaking Lorentz
covariance in the meson sector is annoying in that it requires to boost
composite particles calculated in their rest frame.

\end{document}